\documentclass[12pt]{iopart}
\usepackage{epsfig}
\usepackage{iopams}
\begin{document}
\title{Axial form factor of the nucleon in the perturbative 
chiral quark model}

\author{K Khosonthongkee\dag\ddag,
V E Lyubovitskij\dag , Th Gutsche\dag , Amand Faessler\dag , K
Pumsa-ard\dag , S Cheedket\ddag \,\,and Y Yan\ddag}

\address{\dag \ Institut f\"{u}r Theoretische Physik, Universit\"at
T\"{u}bingen, Auf der Morgenstelle 14, D-72076 T\"{u}bingen,
Germany}
\address{\ddag \ School of Physics, Suranaree University of
Technology, Nakhon Ratchasima 30000, Thailand }

\ead{khoson@tphys.physik.uni-tuebingen.de, 
valeri.lyubovitskij@uni-tuebingen.de,
thomas.gutsche@uni-tuebingen.de, amand.faessler@uni-tuebingen.de, 
pumsa@tphys.physik.uni-tuebingen.de, part@physics2.sut.ac.th,  
yupeng@ccs.sut.ac.th}

\begin{abstract}

We apply the perturbative chiral quark model (PCQM) at one loop to
analyze the axial form factor of the nucleon. This chiral quark model
is based on an effective Lagrangian, where baryons are described by
relativistic valence quarks and a perturbative cloud of
Goldstone bosons as dictated by chiral symmetry.
We apply the formalism to obtain analytical
expressions for the axial form factor of the nucleon, which is given in
terms of fundamental parameters of low-energy pion-nucleon physics
(weak pion decay constant, strong pion-nucleon form factor) and of
only one model parameter (radius of the nucleonic three-quark
core).
\end{abstract}

\submitto{\JPG}

\maketitle

\section{Introduction}

The nucleon axial form factor is of fundamental significance to
weak interaction properties and to the pion-nucleon interaction. Hence
it provides one important test for theories that attempt to describe
the structure of the nucleon. The current status of the
experimental and theoretical understanding of the axial form
factor of the nucleon is reviewed in~\cite{Bernard,Hagiwara}.
Based on the original development of chiral quark models
with a perturbative treatment of the pion 
cloud~\cite{Theberge}-\cite{Gutsche}, we recently extended the 
relativistic quark model suggested in~\cite{Gutsche_Thesis,Gutsche} 
to describe the low-energy properties of
the nucleon~\cite{LyubovitskijC64}-\cite{LyubovitskijB520}.
In references~\cite{LyubovitskijC64}-\cite{Inoue} we
developed the so-called perturbative chiral quark model (PCQM) in
application to baryon properties such as: sigma-term physics,
electromagnetic form factors of the baryon octet, $\pi N$ scattering 
and electromagnetic corrections, strange nucleon form factors,
electromagnetic nucleon-delta transition, etc. In the present
work we follow up on the earlier investigations and employ the
same model in order to study the axial form factor of the nucleon.

\par
The PCQM is based on an effective chiral Lagrangian describing
quarks as relativistic fermions moving in a self-consistent field
(static potential) $V_{\rm eff}(r) = S(r) + \gamma^{0}V(r)$ which 
is described by a sum of a scalar potential $S(r)$ providing 
confinement and the time component of a vector potential 
$\gamma^{0}V(r)$. Obviously, other possible Lorenz structures 
(e.g., pseudoscalar or axial) are excluded by symmetry principles. 
It is known from lattice simulations that a scalar potential should 
be a linearly rising one and a vector potential thought to be 
responsible for short-range fluctuations of the gluon field 
configurations~\cite{Takahashi}. In our study we approximate 
$V_{\rm eff}(r)$ by a relativistic harmonic oscillator potential 
with a quadratic radial dependence~\cite{LyubovitskijC64}  
\begin{eqnarray}\label{V_hop} 
S(r) = M_1 + c_1 r^2\,, \hspace*{.5cm} 
V(r) = M_2 + c_2 r^2\,.
\end{eqnarray}
The model potential defines unperturbed wave functions for the quarks, 
which are subsequently used to calculate baryon properties.
This potential has no di\-re\-ct connection to the underlying physical 
picture and is thought to serve as an approximation of a realistic 
potential. The vector part of the potential is also a pure 
long-ranged potential and is not responsible for the short-range 
fluctuations of gluon fields. In general, we need a vector potential 
to distinguish between quark and antiquark solutions of the Dirac 
equation with an effective potential. Note, that this type of the 
potential was extensively used in chiral potential 
models~\cite{Gutsche},\cite{TegenA307}-\cite{AbbasJ90}. 
A positive feature of this potential is that most 
of the calculations can be done analytically. As was shown in  
Refs.~\cite{Gutsche},\cite{TegenA307}-\cite{AbbasJ90} and 
later on also checked in the PCQM~\cite{LyubovitskijC64}-\cite{Inoue}, 
this effective potential gives a reasonable description of low-energy  
baryon properties and can be treated as a phenomenological  
approximation of the long-ranged potential dictated by QCD.   

\par

Baryons in the PCQM are described 
as bound states of valence quarks supplemented by a cloud of
Goldstone bosons $(\pi,K,\eta)$ as required by chiral symmetry.
The chiral symmetry constraints will in general introduce a nonlinear
meson-quark interaction, but when considering meson as small
fluctuations we restrict the interaction Lagrangian
up to the quadratic term in the meson fields.
With the derived interaction Lagrangian we do our perturbation
theory in the expansion parameter $1/F$ (where $F$ is the pion
leptonic decay constant in the chiral limit). We also treat the
mass term of the current quarks $(\hat{m},m_s)$ as a
perturbation. Dressing the baryonic three-quark core by a cloud of
Goldstone mesons corresponds effectively to the inclusion of sea-quark
contributions. All calculations are performed at one loop or at
order of accuracy $O(1/F^2,\hat{m},m_s)$.

\par 

To be consistent, we use the unified Dirac equation with 
a fixed static potential both for the ground and 
for the excited quark states. In the Appendix we give details of 
the solutions to the Dirac equation for any excited state. 
Inclusion of excited states should be handled consistently. 
First of all, one should guarantee conservation 
of local symmetries (like gauge invariance). Second, excited states 
should be restricted to energies smaller than the typical application 
scale $\Lambda \approx 1$ GeV of low-energy approaches. An alternative 
possibility to suppress the inclusion of higher-order excited states 
is to introduce a meson-quark vertex form factor~\cite{Gutsche,Oset84}. 
Solving the Dirac equation with a relativistic harmonic oscillator 
potential~(\ref{V_hop}) one can show, that the energy shift between 
the first low-lying $1p$ excited states and the $1s$  
ground state is about 200 MeV. The excited states ($1d$ and $2s$) 
lie about $\sim 370$ MeV above the ground state. The $2p$ and the $1f$ 
states are 530 MeV heavier when compared to the ground state. As soon 
as the typical energy of the ground-state 
quark is about 540 MeV\footnote{This value can be deduced from 
a calculation of octet and decouplet baryon spectrum. Similar 
estimates can be found in other chiral quark calculations.}one can 
restrict to the low-lying $1p$, $1d$ and $2s$ 
excited states with energies smaller than the typical scale 
$\Lambda =  1$ GeV. The requirement of convergence of physical 
observables when including excited states is physically not meaningful 
since it takes states with very large energies where the 
phenomenological low-energy approaches break down. 
One of us showed in Ref.~\cite{Gutsche} that the inclusion of excited 
states to the nucleon and $\Delta$ masses can be convergent when 
using a linearly rising confinement potential, e.g., the use of 
potential with a quadratic radial dependence leads to 
a nonsatisfactory convergence. However, this statement is sensitive 
to the quantity one is testing. On the other hand, our approach 
has some different features in comparison to previous ones 
(see, e.g., Ref.~\cite{Gutsche},\cite{TegenA307}-\cite{AbbasJ90}).  
In particular we perform a consistent renormalization procedure 
when we include meson cloud effects. It gives additional 
contributions to physical quantities which were not taken into 
account before. In Ref.~\cite{Pumsa_ard} we demonstrated that 
excited quark states ($1p$, $1d$ and $2s$) can increase the 
contribution of loop diagrams but in comparison to the leading 
order (three-quark core) diagram this effect was of the order of 
$10\%$. This is why we were interested to study these effects for the 
example of the axial form factor. Again, we truncate the set of 
excited states to the $1p$, $1d$ and $2s$ state with energies which 
satisfy the condition ${\cal E} < \Lambda = 1$ GeV. We do not pretend 
that we have a more accurate estimate of the whole tower of excited 
states. The scheme we use is thought to take the excited states into 
account in an average fashion. We showed that the zeroth-order value 
of the axial nucleon charge is not changed much in the presence of 
meson cloud effects in consistency with chiral 
perturbation theory. 
As an extention of previous work, which is dominantly aimed to 
describe the low-energy static properties of baryons, 
we consider the model prediction for the axial charge and for 
completeness also for the axial form factor. No further parameters 
are adjusted in the present work. 

\par
In the present paper we proceed as follows. First, we describe
the basic features of our approach: the underlying effective
Lagrangian, the unperturbed, that is valence quark, result for
the nucleon description together with the choice of parameters
and a brief overview of perturbation theory when including the
meson fields. The full details of the renormalization technique
can be found in reference~\cite{LyubovitskijC64}. In section III
we concentrate on the detailed analysis of the nucleon axial form
factor in our approach. We derive analytical expressions in terms
of fundamental parameters of low-energy pion-nucleon physics (weak
pion decay constant, strong pion-nucleon form factor) and of only
one model parameter (radius of the three-quark core of the
nucleon).  Numerical results in comparison with data are presented
to test the phenomenological implications of the model.
Finally, section IV contains a summary of our major conclusions.

\section{The perturbative chiral quark model}

Following considerations lay out the basic notions of the perturbative
chiral quark model (PCQM), a relativistic quark model suggested
in~\cite{Gutsche_Thesis,Gutsche} and extended
in~\cite{LyubovitskijC64}-\cite{LyubovitskijB520} for the
study of low-energy properties of baryons.
In this model quarks move in an effective static field, represented
by a scalar $S(r)$ and vector $V(r)$ component with
$V_{{\mathrm {eff}}}(r)=S(r)+\gamma^{0}V(r)$ and $r=|\bi{x}|$, providing
phenomenological confinement. The interaction of quarks with Goldstone
bosons is introduced on the basis of the nonlinear
$\sigma$-model~\cite{GellMann16}. 
The PCQM is then defined by the effective, chirally invariant Lagrangian
${\cal L}_{{\mathrm {inv}}}$~\cite{LyubovitskijD63,LyubovitskijB520}
\begin{eqnarray}\label{Lagrangian_inv}
\fl{\cal L}_{{\mathrm{inv}}}(x)=\bar{\psi}(x)\left\{\rmi\not\!\partial
-\gamma^{0}V(r)-S(r)\left[\frac{U+U^{\dag}}{2}+\gamma^{5}
\frac{U-U^{\dag}}{2}\right]\right\}\psi(x)\nonumber\\
+\frac{F^2}{4}\Tr\left[\partial_{\mu}U\,
\partial^{\mu}U^{\dag}\right],
\end{eqnarray}
with an additional mass term for quarks and mesons
\begin{equation}
{\cal L}_{\chi SB}(x)=-\bar{\psi}(x){\cal
M}\psi(x)-\frac{B}{2}\Tr[\hat{\Phi}^{2}{\cal M}],
\end{equation}
which explicitly breaks chiral symmetry. Here $\psi$ is the quark
field; $U = \rme^{\rmi\hat{\Phi}/F}$ is the chiral field;
$\hat{\Phi}$ is the matrix of pseudoscalar mesons (in the
following we restrict to the $SU(2)$ flavor case, that is
$\hat{\Phi}\rightarrow
\hat{\pi}=\boldsymbol{\pi}\cdot\boldsymbol{\tau}$); $F$ = 88 MeV
is the pion decay constant in the chiral limit; ${\cal
M}={\rm diag}\{\hat{m},\hat{m}\}$ is the mass matrix of current quarks
(we restrict to the isospin symmetry limit with
$m_{u}=m_{d}=\hat{m}=7$ MeV) and
$B=-\big<0|\bar{u}u|0\big>/F^{2}=1.4$ GeV is the quark condensate
parameter. We rely on the standard picture of chiral symmetry
breaking and for the mass of pions we use the leading term in
their chiral expansion (i.e. linear in the current quark mass):
$M_{\pi}^{2}= 2\hat{m}B$.
\par
With the unitary chiral rotation 
$\psi\rightarrow\exp\{-\rmi\gamma^{5}\hat{\Phi}/(2F)\}
\psi$~\cite{Thomas81}-\cite{Jennings} 
the La\-gran\-gi\-an~(\ref{Lagrangian_inv}) transforms into a Weinberg-type
form ${\cal L}^{W}$ containing the axial-vector coupling and the
Weinberg-Tomozawa term~\cite{LyubovitskijB520}:
\begin{eqnarray}\label{wt lagrangian}
{\cal L}^{W}(x)&=&{\cal L}_{0}(x)+{\cal L}^{W}_{I}(x)
+o(\boldsymbol{\pi}^{2}),\\
{\cal L}_{0}(x)&=&\bar{\psi}(x)\biggl\{\rmi
\not\!\partial-S(r)-\gamma^{0}V(r)\biggl\}\psi(x)-\frac{1}{2}
\boldsymbol{\pi}(x)\left(\opensquare+M^{2}_{\pi}\right)
\boldsymbol{\pi}(x),
\nonumber\\
{\cal L}^{W}_{I}(x)&=&\frac{1}{2F}\partial_{\mu}\boldsymbol{\pi}(x)
\bar{\psi}(x)\gamma^{\mu}\gamma^{5}\boldsymbol{\tau}\psi(x)-
\frac{\varepsilon_{ijk}}{4F^{2}}\pi_{i}(x)\partial_{\mu}\pi_{j}(x)
\bar{\psi}(x)\gamma^{\mu}\tau_{k}\psi(x),\nonumber
\end{eqnarray}
where ${\cal L}^{W}_{I}(x)$ is the $O(\boldsymbol{\pi}^{2})$ strong 
interaction Lagrangian and $\opensquare =\partial^{\mu}\partial_{\mu}$.

\par
In our calculation we do perturbation theory in the expansion
parameter $1/F$ (where $F$ is the pion leptonic decay constant in
the chiral limit). We also treat the mass term of the current
quarks as a perturbation. Dressing the baryonic three-quark core
by a cloud of Goldstone mesons corresponds effectively to the
inclusion of sea-quark contributions. All calculations are
performed at one loop or at order of accuracy $O(1/F^2,\hat{m})$.
\par
We expand the quark field $\psi$ in the basis of potential
eigenstates as
\begin{equation}
\psi(x)=\sum_{\alpha}b_{\alpha}
u_{\alpha}(\bi{x})\rme^{-\rmi{\cal E}_{\alpha}t}
+\sum_{\beta}d_{\beta}^{\dag}v_{\beta}(\bi{x})
\rme^{\rmi{\cal E}_{\beta}t},
\end{equation}
where the expansion coefficients $b_{\alpha}$ and
$d^{\dag}_{\beta}$ are the corresponding single quark 
anni\-hilation and antiquark creation operators. The set of quark
$\{u_{\alpha}\}$ and antiquark $\{v_{\beta}\}$ wave functions in
orbits $\alpha$ and $\beta$ are solutions of the static Dirac
equation:
\begin{equation}\label{Dirac_eq quark}
[-\rmi\gamma^{0}\boldsymbol{\gamma}\cdot\boldsymbol{\nabla}
+\gamma^{0}S(r)+
V(r)-{\cal E}_{\alpha}]u_{\alpha}(\bi{x})=0,
\end{equation}
where ${\cal E}_{\alpha}$ is the single quark energy.

\par
The unperturbed nucleon state is conventionally set up by the
product of spin-flavor and color quark wave functions, where the
nonrelativistic single quark wave function is replaced by the
relativistic solution $u_{0}(\bi{x})$ in the ground state.

\par
For a given form of the effective potential $V_{\mathrm {eff}}(r)$ the
Dirac equation in equation (\ref{Dirac_eq quark}) can be solved
numerically. Here, for the sake of simplicity, we use a
variational Gaussian ansatz for the quark wave function given by
the analytical form:
\begin{equation}\label{Gaussian_Ansatz}
u_{0}(\bi{x}) \, = \, N_{0} \, \exp\left(-\frac{\bi{x}^{\,
2}}{2R^{2}}\right) \, \left(
\begin{array}{c}
1\\
\rmi \rho \, \frac{\displaystyle{\boldsymbol{\sigma}
\,\bi{x}}}{\displaystyle{R}}\\
\end{array}
\right) \, \chi_s \, \chi_f\, \chi_c \, ,
\end{equation}
where $N_{0}=\left[\pi^{3/2} R^{3}
\left(1+\frac{3}{2}\rho^{2}\right)\right]^{-1/2}$ is a constant
fixed by the normalization condition $\int \rmd^{3}{\bi x} \,
u^{\dagger}_{0}(\bi x) \, u_{0}(\bi x) \equiv 1$ ; $\chi_s$,
$\chi_f$, $\chi_c$ are the spin, flavor and color quark wave
functions, respectively. Our Gaussian ansatz contains two model
parameters: the dimensional parameter $R$ and the dimensionless
parameter $\rho$. The parameter $\rho$ can be related to the axial
coupling constant $g_{A}^{(0)}$ calculated in zeroth-order (or
three-quark core) approximation:
\begin{equation}\
g_{A}^{(0)}=\frac{5}{3} \biggl(1 - \frac{2\rho^{2}} {1+\frac{3}{2}
\rho^{2}} \biggr)=\frac{5}{3}\biggl(\frac{1+2\gamma}{3}\biggl) ,
\end{equation}
where $\gamma=(1-\frac{3}{2}\rho^{2})/(1+\frac{3}{2}\rho^{2})$ is
a relativistic reduction factor. In our calculations we use a
value of $g_A^{(0)}=1.25$ as obtained in the chiral limit of
chiral perturbation theory~\cite{Gasser88}. The parameter $R$
can be physically understood as the mean radius of the three-quark
core and is related to the charge radius of the proton in the
leading-order (or zeroth-order) approximation as
\begin{equation}
\big<r^{2}_{E}\big>^{P}_{LO} \, = \int \rmd^{3}{\bi x}
u^{\dag}_{0}(\bi{x})\bi{x}^{\,2}u_{0}(\bi{x}) \,= \frac{3R^{2}}{2}
\, \frac{1 \, + \, \frac{5}{2} \, \rho^{2}} {1 \, + \, \frac{3}{2}
\, \rho^{2}} \, .
\end{equation}
The parameter $R$ was fixed as $R = 0.6$ fm in our previous 
study of electromagnetic properties of nucleon~\cite{LyubovitskijC64} 
which corresponds to $\big<r^2_E\big>^P_{LO} = 0.6$ fm$^2$. 

\par
The use of the Gaussian ansatz of equation (\ref{Gaussian_Ansatz}) in
its exact form restricts the scalar and the vector part of the
potential to
\begin{eqnarray}
S(r) &=& \frac{1 - 3\rho^{2}}{2\rho R}+\frac{\rho}{2R^{3}}r^{2},\\
V(r) &=& {\cal E}_{0}-\frac{1 + 3\rho^{2}}{2\rho R} +
\frac{\rho}{2R^{3}}r^{2}.
\end{eqnarray}

\par
The above expressions are introduced as a static mean field
potential confinement in PCQM, hence covariance cannot be
fulfilled. As a consequence matrix elements are frame dependent:
both Galilei invariance of the zeroth order baryon wave functions
and Lorentz boost effects, when considering finite momenta
transfers, are neglected. Approximate techniques~\cite{Birse90,Lu98} 
have been developed to account for these 
deficiencies in static potential models. However, these techniques
do not always agree and lead to further ambiguities in model
evaluations. Furthermore, existing Galilean projection techniques
are known to lead to conflict with chiral symmetry 
constrains~\cite{LyubovitskijD63}. In the present manuscript we 
completely neglect the study of these additional model dependent 
effects. We focus on the role of meson loops, which, as shown in 
the content of cloudy bag model~\cite{Lu98}, are not plagued by 
these additional uncertainties.

\par
According to the Gell-Mann and Low theorem we define the expectation
value of an operator $\hat{\Or}$ in the PCQM by
\begin{equation}\label{Gell-Mann}
\big<\hat{\Or}\big> =
^{B}\!\!\big<\phi_{0}|\sum\limits_{n=0}^{\infty}
\frac{\rmi^{n}}{n!}\int \rmd^{4}x_{1}\ldots \int \rmd^{4}x_{n}
T[{\cal L}^{W}_{I}(x_{1})\ldots {\cal L}^{W}_{I}(x_{n})\hat{\Or}]
|\phi_{0}\big>^{B}_{c}.
\end{equation}
The subscript $c$ refers to contributions from connected graphs
only and ${\cal L}^{W}_{I}(x)$ is the pion-quark interaction
Lagrangian as already indicated in equation (\ref{wt lagrangian}).
The superscript $B$ in equation (\ref{Gell-Mann}) indicates that
the matrix elements are projected on the respective baryon states.
The projection of ``one-body" diagrams on the nucleon state refers
to
\begin{equation}
\chi_{f{^\prime}}^{\dag} \chi_{s'}^{\dag}I^{f' f} \, J^{s^{\prime}
s} \, \chi_{f} \chi_{s} \, \stackrel{Proj.}\longrightarrow \,
\big<N\big|\sum\limits_{i=1}^{3} (I \, J)^{(i)}\big|N\big>,
\end{equation}
where the single-particle matrix element of the operators $I$ and
$J$, acting in flavor and spin space, is replaced by the one
embedded in the nucleon state. For ``two-body" diagrams with two
independent quark indices $i$ and $j$ the projection prescription
reads as
\begin{equation}
\fl\chi_{f{^\prime}}^{\dagger} \chi_{s^{\prime}}^{\dagger} \,
I^{f^{\prime} f}_{1} \, J^{s^{\prime} s}_{1} \, \chi_{f} \chi_{s}
\, \otimes \chi_{k^{\prime}}^{\dagger}
\chi_{\sigma^{\prime}}^{\dagger} \, I^{k^{\prime} k}_{2} \,
J^{\sigma^{\prime} \sigma}_{2} \, \chi_{k} \chi_{\sigma}
\stackrel{Proj.}\longrightarrow \, \big<N\big|\sum\limits_{i \,
\neq \, j}^{3} (I_{1} \, J_{1})^{(i)} \otimes (I_{2} \,
J_{2})^{(j)}\big|N\big> .
\end{equation}

\par
We evaluate equation (\ref{Gell-Mann}) using Wick's theorem and
the appropriate propagators. For the quark field we use a Feynman 
propagator for a fermion in a binding potential. The quark
propagator $\rmi G_\psi(x,y)$ is given by
\begin{equation}
\fl \rmi
G_\psi(x,y)=\big<\phi_{0}\big|T\{\psi(x)\bar\psi(y)\}\big|
\phi_{0}\big> \to \sum \limits_{\alpha} u_{\alpha}(\bi{x}) \bar
u_{\alpha}(\bi{y})\exp[-\rmi{\cal
E}_{\alpha}(x_{0}-y_{0})]\theta(x_{0}-y_{0}),
\end{equation}
where we restrict to the quark states propagating forward in time.
The explicit form of the excited quark wave functions are obtained
analytically as given in Appendix. For the mesons we use the free
Feynman propagator for a boson field with
\begin{equation}
\rmi\Delta_{ij}(x-y)=\big<0\big|T\{\Phi_{i}(x)\Phi_{j}(y)\}\big|0\big>
=\delta_{ij}\int\frac{\rmd^{4}k}{(2\pi)^4\rmi}
\frac{\exp[-\rmi k(x-y)]}{M_{\Phi}^{2}-k^{2}-\rmi\epsilon}.
\end{equation}
\par
To redefine our perturbation series up to a given order in terms
of renormalized quantities a set of counterterms $\delta{\cal L}$
has to be introduced in the Lagrangian. Thereby, the counterterms
play a dual role: (i) to maintain the proper definition of
physical parameters, such as nucleon mass and, in particular, the
nucleon charge, and (ii) to effectively reduce the number of
Feynman diagrams to be evaluated. Here we follow the formalism set
out in reference~\cite{LyubovitskijC64}, but where in the present
work intermediate excited quark states are included in the loop
diagrams. In the following we attach the index $"0\,"$ to the
renormalization constants when we restrict to the contribution of
the ground state quark propagator and the superscript $"F\,"$
when excited quark states are included.
\par
First we introduce the renormalized quark field $\psi^{r}(x)$.
It can be expanded in a set of potential eigenstates
which are solutions of the renormalized Dirac equation with
the full renormalized quark mass of
\begin{eqnarray}
\fl \hat{m}^r_{F}=\hat{m}-\frac{3}{\gamma}\biggl(\frac{1}{4\pi
F}\biggl)^{2}\sum_{\alpha}\int_{0}^{\infty} \rmd k\,
k^{2}\,\frac{1}{\omega(k^{2})(\omega(k^{2})+\Delta{\cal E}
_{\alpha})}\nonumber\\
\times \left[F_{I_\alpha}(k)F^{\dag}_{I_\alpha}(k)
-2\omega(k^{2})F_{I_\alpha}(k)F^{\dag}_{II_\alpha}(k)
+\omega^{2}(k^{2})F_{II_\alpha}(k)F^{\dag}_{II_\alpha}(k)\right].
\end{eqnarray}
The expression for the renormalized quark mass includes
self-energy corrections of the pion cloud, where
\begin{eqnarray}
\fl F_{I_\alpha}(k)\equiv N_{0}N_{\alpha}k\Biggl[\int_{0}^{\infty}
\rmd r\,
r^{2}\,(g_{0}(r)g_{\alpha}(r)-f_{0}(r)f_{\alpha}(r))
\int_{\Omega}\rmd\Omega
\,\rme^{\rmi kr\cos\theta}\,C_{\alpha}\,Y_{l_{\alpha}0}(\theta,\phi)
\nonumber\\
-2\,\rmi\frac{\partial}{\partial k}\int_{0}^{\infty} \,\rmd r\,
r\,(f_{0}(r)f_{\alpha}(r))\int_{\Omega}\rmd \Omega\,\cos\theta\,
\rme^{\rmi kr\cos\theta}\,C_{\alpha}\,Y_{l_{\alpha}0}(\theta,\phi)
\Biggl],\\
\fl F_{II_\alpha}(k)\equiv N_{0}N_{\alpha}\frac{\partial}{\partial
k}\int_{0}^{\infty} \rmd r\,
r\,(g_{0}(r)f_{\alpha}(r)-f_{0}(r)g_{\alpha}(r))\int_{\Omega}\rmd\Omega
\,\rme^{\rmi
kr\cos\theta}\,C_{\alpha}\,Y_{l_{\alpha}0}(\theta,\phi),
\end{eqnarray}
with the pion energy $\omega(k^{2})= \sqrt{M^{2}_{\pi}+k^{2}}$;
$k=|\bi{k}|$ is the pion momentum and $\Delta {\cal E}_\alpha =
{\cal E}_\alpha-{\cal E}_0 $ is the excess of the energy of the
quark in state $\alpha$ with respect to the ground state. The
label $\alpha = ( n l_{\alpha} j m )$ characterizes the quark
state (principal quantum number n, non-relativistic orbital
angular momentum $l_{\alpha}$, total angular momentum and
projection j,m). For the Clebsch-Gordan coefficients we use the
notation
$C_{\alpha}\equiv\big{<}l_{\alpha}0
\frac{1}{2}\frac{1}{2}|j\frac{1}{2}\big{>}$
and $Y_{l_{\alpha}\,0}(\theta,\phi)$ is the usual spherical
harmonic. The explicit form of the radial wave functions
$g_{\alpha}(r)$ and $f_{\alpha}(r)$, of the normalization
constants ($N_{\alpha}$) and of the energy difference
($\Delta{\cal E}_{\alpha}$) are given in Appendix.

When restricting the quark propagator to the ground state the
expression above for the renormalized quark mass reduces to
\begin{equation}
\hat{m}^{r}_{0}=\hat{m}-\frac{27}{400\gamma}
\Biggl(\frac{g_{A}^{(0)}}{\pi
F}\Biggl)^{2}\int_{0}^{\infty} \rmd k \,k^{4}\,\frac{F^{2}_{\pi
NN}(k^{2})}{\omega^{2}(k^{2})},
\end{equation}
where $F_{\pi NN}(k^2)$ is the $\pi NN $ form factor normalized to unity
at zero recoil $(k^{2}=0)$:
\begin{equation}
F_{\pi NN}(k^{2})=\exp\Biggl(-\frac{k^{2}R^{2}}{4}\Biggl)\Biggl[
1+\frac{k^{2}R^{2}}{8}\Biggl(1-\frac{5}{3g_{A}^{(0)}}\Biggl)\Biggl].
\end{equation}
\par
In a second step we renormalize the effective Lagrangian including 
a set of counterterms. The renormalized interaction Lagrangian 
${\cal L}^{W}_{I;r}={\cal L}^{W;str}_{I;r}+{\cal L}^{W;em}_{I;r}$ 
contains a part due to the strong interaction,
\begin{equation}
{\cal L}^{W;str}_{I;r}={\cal L}^{W;str}_{I}+\delta{\cal
L}^{W;str},
\end{equation}
and a piece due to the electromagnetic interaction,
\begin{equation}
{\cal L}^{W;em}_{I;r}={\cal L}^{W;em}_{I}+\delta{\cal L}^{W;em}.
\end{equation}
The strong interaction term ${\cal L}^{W;str}_{I}$ is given by
\begin{equation}
\fl{\cal
L}^{W;str}_{I}(x)=\frac{1}{2F}\partial_{\mu}\boldsymbol{\pi}(x)
\bar{\psi^{r}}(x)\gamma^{\mu}\gamma^{5}\boldsymbol{\tau}\psi^{r}(x)
-\frac{\varepsilon_{ijk}}{4F^{2}}\pi_{i}(x)\partial_{\mu}\pi_{j}(x)
\bar{\psi^{r}}(x)\gamma^{\mu}\tau_{k}\psi^{r}(x).
\end{equation}
The interaction of pions and quarks with the electromagnetic field
is described by~\cite{LyubovitskijB520}
\begin{eqnarray}
\hspace*{-.1cm}
\fl {\cal
 L}^{W;em}_{I}(x)=-eA^{em}_{\mu}\bar{\psi}^{r}(x)Q\gamma^{\mu}
\psi^{r}(x)+ \frac{e}{4F^{2}}A^{em}_{\mu}(x)\bar{\psi^{r}}(x)
\gamma^{\mu}\left[\boldsymbol{\pi}^{2}(x)\tau_{3}
 -\boldsymbol{\pi}(x)\boldsymbol{\tau}\pi^{0}(x)\right]
\psi^{r}(x)\nonumber\\
-eA^{em}_{\mu}(x)\varepsilon_{3ij}
\left[\pi_{i}(x)\partial^{\mu}\pi_{j}(x)
-\frac{\pi_{j}(x)}{2F}\bar{\psi}^{r}(x)\gamma^{\mu}\gamma^{5}
\tau_{i}\psi^{r}(x)\right],
\end{eqnarray}
which is generated by minimal substitution with
\begin{equation}
\fl \partial_{\mu}\psi^{r}\rightarrow
D_{\mu}\psi^{r}=\partial_{\mu}\psi^r+\rmi eQA^{em}_{\mu}\psi^r,
\,\,\,\,\,\,\,\,\,\,\partial_{\mu}\pi_{i}\rightarrow
D_{\mu}\pi_{i}=\partial_{\mu}\pi_{i}+e\varepsilon_{3ij}
A^{em}_{\mu}\pi_{j},
\end{equation}
where $Q$ is the quark charge matrix. The set of counterterms,
denoted by $\delta{\cal L}^{W;str}$ and $\delta{\cal L}^{W;em}$
are explained and given in reference~\cite{LyubovitskijC64}.

\par
Now we consider the nucleon charge and prove that the properly
introduced counterterms guarantee charge conservation. Using
Noether's theorem we first derive from the renormalized Lagrangian
the electromagnetic current operator:
\begin{equation}
j^{\mu}_{r}=j^{\mu}_{\psi^{r}}+j^{\mu}_{\pi}+j^{\mu}_{\psi^{r}\pi}+\delta
j^{\mu}_{\psi^{r}}.
\end{equation}
It contains the quark component ($j^{\mu}_{\psi^{r}}$), the charged
pion component ($j^{\mu}_{\pi}$), the quark-pion component
($j^{\mu}_{\psi^{r}\pi}$) and the contribution of the counterterm
($\delta j^{\mu}_{\psi^{r}}$):
\begin{equation}
\eqalign {j^{\mu}_{\psi^{r}}&=\bar{\psi}^{r}
\gamma^{\mu}Q\psi^{r}=\frac{1}{3}\left(2\bar{u}^{r}
\gamma^{\mu}u^{r}-\bar{d}^{r}\gamma^{\mu}d^{r}\right),\nonumber\\
j^{\mu}_{\pi}&=\varepsilon_{3ij} \,
\pi_i\partial^{\mu}\pi_{j} \, = \, \pi^{-}\rmi\partial^{\mu}
\pi^{+}-\pi^{+}\rmi\partial^{\mu}\pi^{-},\nonumber\\
j^{\mu}_{\psi^{r}\pi}&=-\frac{1}{4F^{2}}\bar{\psi}^{r}
\gamma^{\mu}\left(\boldsymbol{\pi}^{2}\tau_{3}
-\boldsymbol{\pi}\,
\boldsymbol{\tau}\pi^{0}\right)\psi^{r}-\varepsilon_{3ij}
\frac{\pi_j}{2F}\bar{\psi^{r}}
\gamma^{\mu}\gamma^{5}\tau_{i}\psi^{r},\nonumber\\
\delta
j^{\mu}_{\psi^{r}}&=\bar{\psi}^{r}(\hat{Z}-1)\gamma^{\mu}Q\psi^{r}}
\end{equation}
Here, $\hat{Z}$ is the renormalization constant determined by the
nucleon charge conservation condition~\cite{LyubovitskijC64}. In
the one-loop approximation following diagrams contribute to the
nucleon charge [Figs.1a-1f]: the three-quark diagram [Fig.1a] 
with an insertion of the quark current 
$j^{\mu}_{\psi^{r}}$, the three-quark diagram [Fig.1b] with
the counterterm $\delta j^{\mu}_{\psi^{r}}$ (three-quark
counterterm diagram), the self-energy [Figs.1c and 1d], the
vertex correction diagram [Fig.1e] with the quark current
$j^{\mu}_{\psi^{r}}$, and finally the meson-cloud 
diagram [Fig.1f] generated by the pion current $j^{\mu}_{\pi}$.

The analytical expression for the full renormalization constant,
$\hat{Z}^{F}$ is
\begin{eqnarray}\label{ZF}
\fl \hat{Z}^{F}=1-\frac{3}{(4\pi F)^{2}}
\sum_{\alpha}\int_{0}^{\infty} \rmd k\,
k^{2}\,\frac{1}{\omega(k^{2})(\omega(k^{2})+\Delta{\cal E}
_{\alpha})^{2}}\nonumber\\
\times\left[F_{I_\alpha}(k)
F^{\dag}_{I_\alpha}(k)-2\,\omega(k^{2})
F_{I_\alpha}(k)F^{\dag}_{II_\alpha}(k)
+\omega^{2}(k^{2})F_{II_\alpha}(k)F^{\dag}_{II_\alpha}(k)\right].
\end{eqnarray}
When restricting intermediate quark states to the ground state
equation (\ref{ZF}) yields the result
\begin{equation}
\hat{Z}^{0}=1-\frac{27}{400}\Biggl(\frac{g_{A}^{(0)}}{\pi
F}\Biggl)^{2} \int_{0}^{\infty} \rmd k\,k^4\, \frac{F^{2}_{\pi
NN}(k^2)}{\omega^3(k^2)}.
\end{equation}
We obtain a value of $\hat{Z}^{0}=0.9$~\cite{LyubovitskijC64}
for our set of parameters. Inclusion of the excited quark states changes
the value of the renormalization constant to a value of
$\hat{Z}^{F}=0.7$.

\section{The axial form factor of nucleon}

For the present purposes we have to construct the partially
conserved axial-vector current $A^{\mu}_i$:
\begin{eqnarray}
\fl A^{\mu}_{i}=F\partial^{\mu}\pi_{i}\,+\,\bar{\psi^{r}}
\gamma^{\mu}\gamma^{5} \frac{\tau_{i}}{2}\psi^{r} -
\frac{\varepsilon_{ijk}}{2F}\bar{\psi}^{r}\gamma^{\mu}\tau_{j}
\psi^{r}\pi_{k} \,+\,\frac{1}{4 F^2} \,
\bar{\psi^{r}}\gamma^{\mu}\gamma^{5}
\left(\boldsymbol{\pi}\,\boldsymbol{\tau} \, \pi_{i} -
\boldsymbol{\pi}^{2} \, \tau_{i}\right)\psi^{r} \\
+\bar{\psi^{r}}  (\hat{Z}-1) \gamma^{\mu}\gamma^{5}
\frac{\tau_{i}}{2} \psi^{r} \, + \, o(\boldsymbol{\pi}^{2}) \, .
\nonumber
\end{eqnarray}
The axial form factor $G_A(Q^2)$ of the nucleon is defined by
the matrix element of the $i=3$ isospin component and the spatial
part of the axial vector current evaluated for nucleon states.
In the Breit frame $G_A(Q^2)$ is set up as~\cite{TegenA314}:
\begin{equation}
\left< N_{s'}\left( \frac{\bi{q}}{2}\right)\biggl|\int
\rmd^{3}\bi{x}\,\rme^{\rmi
\bi{q}\,\bi{x}}\bi{A}_{3}(x)\biggl|N_{s}\left(-\frac{\bi{q}}{2}
\right)\right>=\chi^{\dag}_{N_{s'}}\boldsymbol{\sigma}_{N}
\frac{\tau_N^3}{2}
\chi_{N_{s}}G_A(Q^2),
\end{equation}
with $Q^{2}=-q^{2}$. Here, $\chi_{N_{s}}$ and $\chi_{N_{s'}}$ are
the nucleon spin wave functions in the initial and final states;
$\boldsymbol{\sigma}_{N}$ is the nucleon spin matrix and
$\tau_N^3$ is the third component of the isospin matrix of the
nucleon. At zero recoil $(Q^2=0)$ the axial form factor satisfies
the condition $G_A(0)=g_A$, where $g_A$ is axial charge
of the nucleon.

\par
In the PCQM the axial form factor of the nucleon up to one loop
corrections is initially given by
\begin{eqnarray}\label{axial}
\fl\chi^{\dag}_{N_{s'}}\boldsymbol{\sigma}_{N}\frac{\tau_N^3}{2}
\chi_{N_{s}}G_A(Q^2) = ^N\!\!\big<\phi_{0}| \,
\sum\limits_{n=0}^{2} \frac{\rmi^{n}}{n!} \, \int \, \delta(t) \,
\rmd^{4}x 
\,\rmd^{4}x_{1}\, \ldots \, 
\rmd^{4}x_{n} \, \rme^{-\rmi qx} \,\nonumber\\
\times T[{\cal L}^{W;str}_{I}(x_{1}) \, \ldots \, {\cal
L}^{W;str}_{I}(x_{n}) \, \bi{A}_{3}(x)] \, |\phi_{0}\big>_{c}^{N},
\end{eqnarray}
with the interaction term
\begin{equation}
\fl{\cal L}^{W;str}_{I}(x)=\frac{1}{2F}\partial_{\mu}\boldsymbol{\pi}(x)
\bar{\psi^{r}}(x)\gamma^{\mu}\gamma^{5}\boldsymbol{\tau}\psi^{r}(x)-
\frac{\varepsilon_{ijk}}{4F^{2}}\pi_{i}(x)\partial_{\mu}\pi_{j}(x)
\bar{\psi^{r}}(x)\gamma^{\mu}\tau_{k}\psi^{r}(x),
\end{equation}
where the superscript $r$ refers to renormalized quantities.
For the final calculation we include a set of excited states
up to $2\hbar \omega$ in the quark propagator: the first p-states
($1p_{1/2}$ and $1p_{3/2}$ in the non-relativistic spectroscopic 
notation) and the second excited states ($1d_{3/2},\,1d_{5/2}$ and
$2s_{1/2}$). In other words we include the excited states whose 
energies satisfy to the restriction ${\cal E} < \Lambda = 1$ GeV 
(see discussion in the Sect.1). The diagrams to be evaluated are 
shown in Fig.2. 

Next, we present the analytical expressions for the axial form factor
of the nucleon obtained in the PCQM. We start with the simplest
case, where the quark propagator is restricted to the ground state
contribution. The axial form factor of the nucleon is a sum of
terms arising from different diagrams: the three-quarks diagram
[Fig.2a], the counterterm [Fig.2b], the self-energy
diagrams [Figs.2c and 2d], the exchange diagram [Fig.2e] 
and the vertex-correction diagram [Fig.2f]. Other
possible diagrams at one loop are compensated by the counterterm.
The corresponding analytical expressions for the relevant diagrams
are given in the following.
\par 
(a) For the three-quark diagram $(3q)$ [Fig.2a] we obtain:
\begin{equation}
G_{A}(Q^2)\bigg|_{3q}=G_A(Q^{2})\bigg|_{3q}^{LO} + 
G_{A}(Q^2)\bigg|_{3q}^{NLO},
\end{equation}
where $G_{A}(Q^{2})|_{3q}^{LO}$ is the leading-order term (LO)
evaluated with the unperturbed quark wave function (w.f.)
$u_{0}(\bi{x})$; $G_{A}(Q^{2})|_{3q}^{NLO}$ is a correction due to
the renormalization of the quark w.f. $u_{0}(\bi{x})\rightarrow
u_{0}^{r}(\bi{x};\hat{m}^{r})$ which is referred to as next-to-leading
order (NLO):
\begin{eqnarray}
\fl G_{A}(Q^{2})\bigg|_{3q}^{LO}\,\,\,=g_{A}^{(0)}F_{\pi NN}(Q^{2})\\
\fl G_{A}(Q^{2})\bigg|_{3q}^{NLO}=\frac{3}{2}\,\hat{m}^{r}_{0}
\frac{\rho\,R}{\left(1+\frac{3}{2}\rho^{2}\right)^{2}}
\Biggl\{\left(1+\frac{9}{2}
\rho^{2}\right)G_{A}(Q^{2})\bigg|_{3q}^{LO}\nonumber\\
-\frac{5}{72}\biggl[12\left(2-3\rho^{2}\right)
-4\left(1+5\rho^{2}\right)Q^{2}
R^{2}+\rho^{2}Q^{4}R^{4}\biggl]
\exp\left(-\frac{Q^{2}R^{2}}{4}\right)\Biggr\}.
\end{eqnarray}
The modified quark w.f. $u_{0}^{r}(\bi{x};\hat{m}^{r})$
is given by~\cite{LyubovitskijC64}
\begin{equation}
u_{0}^{r}(\bi{x};\hat{m}^{r})=u_{0}(\bi{x})+\delta
u_{0}^{r}(\bi{x};\hat{m}^{r}),
\end{equation}
where
\begin{equation}
\delta
u_{0}^{r}(\bi{x};\hat{m}^{r})=\frac{\hat{m}^{r}}{2}\frac{\rho
R}{1+\frac{3}{2}\rho^{2}}\left(\frac{\frac{1}{2}+\frac{21}{4}\rho^{2}}
{1+\frac{3}{2}\rho^{2}}-\frac{\bi{x}^{2}}{R^{2}}+\gamma^{0}\right)
u_{0}(\bi{x}).
\end{equation}
\par
(b) The three-quark counterterm (CT) [Fig.2b] results in the
expression:
\begin{equation}
G_{A}(Q^{2})\bigg|_{CT}\equiv(\hat{Z}^{0}-1)G_{A}(Q^{2})
\bigg|_{3q}^{LO}.
\end{equation}
\par
(c) The self-energy diagram I (SE;I) [Fig.2c] yields:
\begin{equation}
G_{A}(Q^{2})\bigg|_{SE;I}=8 \,  g_{A}^{(0)} \, \frac{\rho\,R}
{(2+3\rho^{2})}\left(\frac{1}{2\pi
F}\right)^{2}\int^{\infty}_{0}\!\! \rmd k\,k^{4}\frac{F_{\pi
NN}(k^{2})}{\omega^{2}(k^{2})}\,{\cal D}(k,Q^{2}),
\end{equation}
where
\begin{eqnarray}
\fl {\cal
D}(k,Q^{2})\equiv\exp\left(-\frac{(k+\sqrt{Q^2})^{2}R^{2}}{4}\right)
\left(\frac{1}{k^{2}Q^{2}R^{4}}\right)^{3/2}\nonumber\\
\times \left[2+k\sqrt{Q^{2}}R^{2}+\exp\left(k\sqrt{Q^{2}}R^{2}\right)
\left(-2+k\sqrt{Q^{2}}R^{2}\right)\right].
\end{eqnarray}
\par
(d) For the self-energy diagram II (SE;II) [Fig.2d] we also
obtain:
\begin{equation}
G_{A}(Q^2)\bigg|_{SE;II}=G_{A}(Q^2)\bigg|_{SE;I}.
\end{equation}
\par
(e) For the exchange diagram (EX) [Fig.2e] we get:
\begin{equation}
G_{A}(Q^{2})\bigg|_{EX}=\frac{96}{5} \, g_{A}^{(0)} \,
\frac{\rho\,R}{(2+3\rho^{2})}\Biggl(\frac{1}{2\pi
F}\Biggl)^{2}\int^{\infty}_{0}\!\!\rmd k\,k^{4}\frac{F_{\pi
NN}(k^2)}{\omega^{2}(k^{2})}\,{\cal D}(k,Q^{2}).
\end{equation}
\par
(f) The vertex-correction diagram (VC) [Fig.2f] gives
the contribution:
\begin{equation}
G_{A}(Q^{2})\bigg|_{VC}=\frac{3}{100}\left(g_{A}^{(0)}\right)^{3}
\left(\frac{1}{2\pi F}\right)^{2}F_{\pi NN}(Q^{2})\int_{0}^{\infty}
\rmd k\,k^{4} \, \frac{F^{2}_{\pi NN}(k^{2})}{{\omega}^{3}(k^{2})}.
\end{equation}
\par
In the next step we extend the formalism by also including
excited states in the quark propagator.
The leading-order expression of the three-quark diagram and
the exchange term remain the same.
In turn the following contributions must be extended.
\par
(a) In the three-quark NLO expression the appropriate renormalized 
mass has to be inserted with
\begin{eqnarray}\label{3q-NLO total}
\fl G_{A}(Q^{2})\bigg|_{3q}^{NLO}=\frac{3}{2}\,\hat{m}^{r}_{F}
\frac{\rho\,R}{\left(1+\frac{3}{2}\rho^{2}\right)^{2}}
\Biggl\{\left(1+\frac{9}{2}
\rho^{2}\right)G_{A}(Q^{2})\bigg|_{3q}^{LO}\nonumber\\
-\frac{5}{72}\bigg[12\left(2-3\rho^{2}\right)-4\left(1+5\rho^{2}\right)
Q^{2}R^{2}+\rho^{2}Q^{4}R^{4}\bigg]\exp\left(-\frac{Q^{2}R^{2}}{4}
\right)\Biggr\}.
\end{eqnarray}
\par
(b) For the three-quark counterterm (CT) the
renormalization constant has to be replaced accordingly
\begin{equation}
G_{A}(Q^{2})\bigg|_{CT}=(\hat{Z}^{F}-1)G_A(Q^2)\bigg|_{3q}^{LO}.
\end{equation}
\par
(c) For the self-energy diagram I (SE;I) we obtain the full
expression
\begin{eqnarray}
\fl G_{A}(Q^{2})\bigg|_{SE;I}=\frac{10}{3}\left(\frac{1}{4\pi
F}\right)^{2} \sum_{\alpha}\int_{0}^{\infty}\rmd
k\,k^{2}\,\frac{\omega(k^{2})
F_{II_{\alpha}}(k)-F_{I_{\alpha}}(k)}{\omega(k^{2})(\omega(k^{2})+
\Delta{\cal E}_{\alpha})}\nonumber\\
\times\int_{-1}^{1}\,\rmd
x\,\frac{2k(1-x^{2})F_{III_{\alpha}}(k_{-})+
\left(\sqrt{Q^{2}}x+(1-2x^{2})k\right)
F_{IV_{\alpha}}(k_{-})}{\sqrt{k_{-}^{2}}},
\end{eqnarray}
where
\begin{eqnarray}
\fl F_{III_{\alpha}}(k_{-})\equiv N_{0}
N_{\alpha}\frac{\partial}{\partial k_{-}}\int_{0}^{\infty}\rmd
r\,r\left(g_{0}(r)f_{\alpha}(r)\right)\int_{\Omega}\rmd\Omega\,
\e^{\rmi k_{-}\,r\,\cos\theta}C_{\alpha}Y_{l_{\alpha}0}(\theta,\phi),
\nonumber\\
\fl F_{IV_\alpha}(k_{-})\equiv N_{0}
N_{\alpha}\frac{\partial}{\partial k_{-}}\int_{0}^{\infty}\rmd
r\,r\left(f_{0}(r)g_{\alpha}(r)-g_{0}(r)f_{\alpha}(r)\right)\nonumber\\
\times \int_{\Omega}\rmd\Omega\, \e^{\rmi
k_{-}\,r\,\cos\theta}C_{\alpha}Y_{l_{\alpha}0}(\theta,\phi),\nonumber\\
\fl k^{2}_{\pm}\equiv k^{2}+Q^{2}\pm2k\sqrt{Q^{2}}x.
\end{eqnarray}
\par
(d) For the self-energy diagram II (SE;II) we get
\begin{eqnarray}
\fl G_{A}(Q^{2})\bigg|_{SE;II}=\frac{10}{3}\left(\frac{1}{4\pi
F}\right)^{2}\sum_{\alpha}\int_{0}^{\infty}\rmd
k\,k^{2}\,\frac{\omega(k^{2})F_{II_{\alpha}}^{\dag}(k)
-F_{I_{\alpha}}^{\dag}(k)}
{\omega(k^{2})(\omega(k^{2})+\Delta {\cal E}_{\alpha})}\nonumber\\
\times\int_{-1}^{1}\,\rmd x\,\frac{2k(1-x^{2})F_{V_{\alpha}}
(k_{+})-(\sqrt{Q^{2}}x+k)F_{IV_{\alpha}}(k_{+})}{\sqrt{k_{+}^{2}}},
\end{eqnarray}
where
\begin{equation}
\fl F_{V_{\alpha}}(k_{+})\equiv N_{0}
N_{\alpha}\frac{\partial}{\partial k_{+}}\int_{0}^{\infty}\rmd
r\,r\left(f_{0}(r)g_{\alpha}(r)\right)\int_{\Omega}\rmd\Omega
\e^{\rmi
k_{+}\,r\,\cos\theta}C_{\alpha}Y_{l_{\alpha}0}(\theta,\phi).
\end{equation}
\par
(e) For the vertex-correction diagram (VC) inclusion of
excited states results in
\begin{eqnarray}
\fl G_{A}(Q^{2})\bigg|_{VC}=\sum_{\alpha\,\beta}\frac{5}{9}
\frac{{\cal F}_{\alpha,\beta}(Q^{2})}{(4\pi F)^{2}}
\int_{0}^{\infty}\rmd k\,k^{2}\,\left[\frac{1}{\omega(k^{2})
(\omega(k^{2}) +\Delta{\cal
E}_{\alpha})\left(\omega(k^{2})+\Delta{\cal
E}_{\beta}\right)}\right]\nonumber\\
\times\biggl[\omega^{2}(k^{2})\left(F_{II_{\alpha}}(k)
F_{II_{\beta}}^{\dag}(k)\right)
-\omega(k^{2})\left(F_{II_{\alpha}}(k)F_{I_{\beta}}^{\dag}(k)
+F_{I_{\alpha}}(k)F_{II_{\beta}}^{\dag}(k)\right)\nonumber\\
+\left(F_{I_{\alpha}}(k)F_{I_{\beta}}^{\dag}(k)\right)\biggl],
\end{eqnarray}
where
\begin{eqnarray}
\fl{\cal F_{\alpha,\beta}}(Q^{2})\equiv
N_{\alpha}N_{\beta}\int_{0}^{\infty}\,\rmd r\,r^{2}{\cal
\bigg(A_{\alpha,\beta}}(r)+2{\cal B_{\alpha,\beta}}(r)\bigg),\\
\fl{\cal A}_{\alpha,\beta}(r)\equiv
\left(g_{\alpha}(r)g_{\beta}(r)-f_{\alpha}(r)f_{\beta}(r)\right)
\int_{\Omega}\rmd\Omega\,\exp\left(\rmi\sqrt{Q^{2}}r\cos\theta\right)\,
{\cal C}_{\alpha\beta;1}(\theta,\phi),\\
\fl{\cal B}_{\alpha,\beta}(r)\equiv
f_{\alpha}(r)f_{\beta}(r)\int_{\Omega}\rmd\Omega\,
\exp\left(\rmi\sqrt{Q^{2}}r\cos\theta\right)\nonumber\\
\times\biggl[\cos^{2}\theta\, {\cal
C}_{\alpha\beta;1}(\theta,\phi)+\sin\theta\,\cos\theta \,
{\cal C}_{\alpha\beta;2} (\theta,\phi)\biggl],\\
\fl{\cal C}_{\alpha\beta;1}(\theta,\phi)\equiv C_{\alpha}C_{\beta}
Y_{l_{\alpha}0}(\theta,\phi)Y_{l_{\beta}0}(\theta,\phi)
-D_{\alpha}D_{\beta}Y_{l_{\alpha}1}^{*}(\theta,\phi)Y_{l_{\beta}1}
(\theta,\phi),\\
\fl{\cal C}_{\alpha\beta;2}(\theta,\phi)\equiv C_{\alpha}D_{\beta}
Y_{l_{\alpha}0}(\theta,\phi)Y_{l_{\beta}1}(\theta,\phi)\rme^{-\rmi\phi}
+D_{\alpha}C_{\beta}Y_{l_{\alpha}1}^{*}(\theta,\phi)Y_{l_{\beta}0}
(\theta,\phi)\rme^{\rmi\phi},
\end{eqnarray}
where
$D_{\alpha}=\big{<}l_{\alpha}1\frac{1}{2} \,
{\rm -}\frac{1}{2}|j\frac{1}{2}\big{>}$,
$l_{\alpha}$ and $l_{\beta}$ are the orbital quantum numbers of
the intermediate states $\alpha$ and $\beta$, respectively.

\par

The $Q^{2}$-dependence (up to $0.4\,{\rm GeV}^{2}$) of the axial
form factor of the nucleon are shown in Figs.3, 4 and 5, the
description of each figure is given below. Due to the lack of
covariance, the form factor can be expected to be reasonable up to
$Q^{2}<\bi{p}^{2}=0.4\,\rm{GeV}^{2}$, where $\bi{p}$ is the
typical three-momentum transfer which defines the region where
relativistic effect $\leq\,10\%$ or where the following inequality
$\bi{p}^{2}/\left(4m^{2}_{N}\right)<0.1$ is fulfilled.

The first result for the $Q^{2}$-dependence of the axial form 
factor $G_{A}(Q^{2})$ of the nucleon is indicated in Fig.3.  
The numerical values are obtained, when truncating the quark 
propagator to the ground state or equivalently to the intermediate
nucleon and delta baryon states in loop diagrams. Thereby, we also
give the individual contributions of the different diagrams of 
Fig.2, which add up coherently. The leading order three-quark
diagram dominates the result for the axial form factor, whereas
pion cloud corrections add about $20\%$ of the total result. Here,
both the exchange and self-energy terms give the largest, positive
contribution.

In a next step we include the intermediate excited quark states
with quantum numbers $1p_{1/2},1p_{3/2},1d_{3/2},1d_{5/2}$ and
$2s_{1/2}$ in the propagator. The resulting effect on
$G_{A}(Q^{2})$ is given in Fig.4. We explicitly indicate the
additional terms, which are solely due to the contribution of
these excited states. The previous result, where the quark
propagator is restricted to the ground state, is contained in the
curve denoted by Total(GS). The inclusion of the intermediate
excited states tends to induce a cancellation of the original pion
cloud corrections generated for the case of the ground state quark
propagator, thereby regaining approximately the tree level result.
In Fig.5 we give for completeness the full result for
$G_{A}(Q^{2})$ including excited states in comparison
with experimental data and with the dipole fit using an axial mass
of $M_{A}=1.069$ GeV and normalized to $G_{A}(0) = 1.267$ 
at zero recoil. The model clearly underestimates the finite
$Q^2$-behavior, but it should be noted that a similar effect
occurs in the discussion of the electromagnetic form factors of
the nucleon~\cite{LyubovitskijC64}. The stiffness of the form
factors can be traced to the Gaussian ansatz of the single quark
wave functions and can also be improved when in addition resorting
to a fully covariant description of the valence quark content of
the nucleon~\cite{Ivanov1996}. Hence the applicability of the PCQM
is mostly for static quantities and low $Q^2$ observables of baryons.
\par
For the comparison with the data near $Q^{2}=0$ we first turn to the
results for the axial charge, $g_{A}$. In Table 1 we list the
numerical values for the complete set of Feynman diagrams (Fig.2),
again indicating separately the contributions of ground and
excited states in the quark propagator. 

The prediction for the axial charge including loop corrections are 
relevant for several reasons: 
1) the tree level result for $g_A$ was previously adjusted to fix one 
($\rho$) of the parameters. Since loop corrections essentially do not 
change this results, the previous model predictions remain meaningful; 
2) the predicted small loop corrections to $g_A$ are consistent with 
similar results in chiral perturbation theory;  
3) the generic role of excited states in loop diagrams are rather 
relevant in understanding the nucleon properties. This role was 
already exemplified in the case of $N-\Delta$ 
transition~\cite{Pumsa_ard} and sigma-terms~\cite{Inoue} and 
again is demonstrated in the case of $g_A$.

\par
For low-momentum transfers, that is $Q^2\leq 1$ GeV$^2$, the axial form
factor can be represented by a dipole fit
\begin{eqnarray}
G_{A}(Q^{2})=\frac{G_{A}(0)}{(1+\frac{Q^{2}}{M_{A}^{2}})^{2}},
\end{eqnarray}
in terms of one adjustable parameter $M_A$, the axial mass (or
sometimes dipole mass). Therefore, the axial
radius can be expressed in terms of the axial mass with:
\begin{eqnarray}
\big<r^{2}_{A}\big>=-6\frac{1}{G_{A}(0)}\frac{dG_{A}(Q^{2})}{dQ^{2}}
\bigg|_{Q^{2}=0}\,\,=\frac{12}{M_{A}^{2}}.
\end{eqnarray}
A comparison of the experimentally deduced values for the axial mass and
the axial radius with our model results is given in Table 2.

\section{Summary and Conclusions}
In summary, we have evaluated the axial form factor of the nucleon
and, more important, its low $Q^2$ limits, such as the axial
charge and the axial radius using a perturbative chiral quark
model as based on an effective chiral Lagrangian. Since the PCQM
is a static model, Lorentz covariance cannot be fulfilled.
Approximate techniques to account for Galilei invariance and
Lorentz boost effects were shown to change the tree level results
by about $10\%$. Higher order, that is loop contributions, are
less sensitive to these correction. The derived quantities
contain, in consistency with previous works, only one model
parameter \emph{R}, which is related to the radius of the
three-quark core, and are otherwise expressed in terms of
fundamental parameters of low energy hadron physics: weak pion
decay constant, and set of QCD parameters. In addition, another
parameter ($\rho$ of equation (6)), which is related to the
amplitude of the small component of the single quark wave
function, was originally set up to reproduce the value for the
axial charge in the chiral limit with
$g^{(0)}_A=1.25$~\cite{Gasser88}. Predictions are given for 
the fixed values of model parameters $\rho$ and $R$ 
in consistency with previous investigations. 
In particular, our result for the axial charge,
$g_{A}=1.19$, is in reasonable agreement with the
central value of data: $g_{A}=1.267\pm0.003$. Thereby,
contributions of excited quark states in loop diagrams play a
considerable role in order to generate a small correction to the
tree level result, which is required to account for the data
point. This result, obtained in the context of the PCQM, is rather
encouraging. Minor pion cloud corrections to the tree level result
of $g_A$ justify in turn the appropriate choice for $\rho$ or for
$g^{(0)}_A$ used in previous works. Also, recent calculations of
the axial charge up to order $p^4$ in chiral perturbation
theory~\cite{GA_CHPT1,GA_CHPT2} imply rather large $p^3$
corrections leading to rather large uncertainties when going to
the next order in the chiral expansion. Our model result can
naturally explain the small correction to the one obtained in the
chiral limit, but only when going beyond nucleon and delta states
in the loop diagrams.

\section*{Acknowledgments}
This work was supported by the Deutsche Forschungsgemeinschaft
(DFG Nos.FA67/25-3, GRK 683). K. K., S. C. and Y. Y. acknowledge
the support of Thailand Research Fund (TRF, Grant No.RGJ
PHD/00187/2541, PHD/00165/2541) and the Deutscher Akademischer
Austauschdienst (DAAD, Grant No. A/01/19908). K. P. thanks the
Development and Promotion of Science and Technology Talent Project
(DPST), Thailand for financial support.

\section*{Appendix: Solution of Dirac equation 
for the effective potential}
In this appendix we indicate the solutions to the Dirac equation
with the effective potential $V_{\rm{eff}}(r)=S(r)+\gamma^{0}V(r)$
, with $r=|\bi{x}|$. The particular choice of scalar $S(r)$ and
the time-like vector $V(r)$ parts are given by
\begin{eqnarray}
S(r)=\frac{1-3\rho^{2}}{2\rho R}+\frac{\rho}{2R^{3}}\,\,r^{2},\\
V(r)={\cal E}_{0}-\frac{1 + 3\rho^{2}}{2\rho
R}+\frac{\rho}{2R^{3}}\,\,r^{2}.
\end{eqnarray}
The quark wave function $u_{\alpha}(\bi{x})$ in state $\alpha$ and
eigenenergy ${\cal E}_{\alpha}$ with the specific choice of
$V_{\rm{eff}}$  satisfies the Dirac equation
\begin{eqnarray}\label{Dirac_eq quark Appendix}
[-\rmi\gamma^{0}\boldsymbol{\gamma}\cdot\boldsymbol{\nabla}
+\gamma^{0}S(r)+V(r)-{\cal E}_{\alpha}]u_{\alpha}(\bi{x})=0.
\end{eqnarray}
The solution of the Dirac spinor $u_{\alpha}(\bi{x})$ to
(\ref{Dirac_eq quark Appendix}) can be written in the analytical
form~\cite{TegenA307}:
\begin{equation}
u_{\alpha}(\bi{x}) \, = \, N_{\alpha} \, \left(
\begin{array}{c}
g_{\alpha}(r)\\
\rmi\boldsymbol{\sigma}\cdot \hat{\bi{r}}f_{\alpha}(r)\\
\end{array}
\right) \,{\cal Y}_{\alpha}(\hat{\bi{r}})\chi_f\, \chi_c\,.
\end{equation}
For the particular choice of potential the radial functions
$g_{\alpha}(r)$ and $f_{\alpha}(r)$ satisfy the form
\begin{eqnarray}
g_{\alpha}(r)=\left(\frac{r}{R_{\alpha}}\right)^{l}L^{l+1/2}_{n-1}
\left(\frac{r^{2}}{R^{2}_{\alpha}}\right)
\rme^{-\frac{r^{2}}{2R^{2}_{\alpha}}},
\end{eqnarray}
where for the $j=l+\frac{1}{2}$
\begin{eqnarray}
f_{\alpha}(r)=\rho_{\alpha}\left(\frac{r}{R_{\alpha}}\right)^{l+1}
\left[L^{l+3/2}_{n-1}\left(\frac{r^{2}}{R^{2}_{\alpha}}\right)
+L^{l+3/2}_{n-2}\left(\frac{r^{2}}{R^{2}_{\alpha}}\right)\right]
\rme^{-\frac{r^{2}}{2R^{2}_{\alpha}}},
\end{eqnarray}
and for $j=l-\frac{1}{2}$
\begin{eqnarray}
\fl
f_{\alpha}(r)=-\rho_{\alpha}\left(\frac{r}{R_{\alpha}}\right)^{l-1}
\left[\left(n+l-\frac{1}{2}\right)L^{l-1/2}_{n-1}
\left(\frac{r^{2}}{R^{2}_{\alpha}}\right)+n\,L^{l-1/2}_{n}
\left(\frac{r^{2}}{R^{2}_{\alpha}}\right)\right]
\rme^{-\frac{r^{2}}{2R^{2}_{\alpha}}}.
\end{eqnarray}
The label $\alpha=(nljm_{j})$ characterizes the state with
principle quantum number $n=1,2,3,...,$ orbital angular momentum
$l$, total angular momentum $j=l\pm\frac{1}{2}$ and projection
$m_{j}$. Due to the quadratic nature of the potential the radial
wave functions contain the associated Laguerre polynomials
$L^{k}_{n}(x)$ with
\begin{eqnarray}
L^{k}_{n}(x)=\sum^{n}_{m=0}(-1)^m
\frac{(n+k)!}{(n-m)!(k+m)!m!}x^{m}.
\end{eqnarray}
The angular dependence, ${\cal Y}_{\alpha}(\hat{\bi{r}})\equiv
{\cal Y}_{ljm_{j}}(\hat{\bi{r}})$, is defined by
\begin{eqnarray}
{\cal Y}_{ljm_{j}}(\hat{\bi{r}})=\sum_{m_{l},m_{s}}\left<
l\,m_{l}\frac{1}{2}\,m_{s}\Big|j\,m_{j}\right>Y_{lm_{l}}(\hat{\bi{r}})
\chi_{\frac{1}{2}m_{s}}
\end{eqnarray}
where $Y_{lm_{l}}(\hat{\bi{r}})$ is the usual spherical harmonic.
Flavor and color part of the Dirac spinor are represented by
$\chi_{f}$ and $\chi_{c}$, respectively.

The two coefficients $R_{\alpha}$ and $\rho_{\alpha}$ in the
redial function of state $\alpha$ are of the form
\begin{eqnarray}
R_{\alpha}&=&R(1+\Delta{\cal E}_{\alpha}\rho R)^{-1/4},\\
\rho_{\alpha}&=&\rho\left(\frac{R_{\alpha}}{R}\right)^{3}
\end{eqnarray}
and are related to the Gaussian parameter $\rho, R$ of equation
(\ref{Gaussian_Ansatz}). The quantity $\Delta{\cal
E}_{\alpha}={\cal E}_{\alpha}-{\cal E}_{0}$ is the excess of the
energy of the quark state $\alpha$ with respect to the ground
state. $\Delta {\cal E}_{\alpha}$ depends on the quantum number
$n$ and $l$ and is related to the parameters $\rho$ and $R$ by
\begin{eqnarray}
\left(\Delta {\cal
E}_{\alpha}+\frac{3\rho}{R}\right)^{2}\left(\Delta {\cal
E}_{\alpha}+\frac{1}{\rho
R}\right)=\frac{\rho}{R^{3}}(4n+2l-1)^{2}.
\end{eqnarray}

The normalization constant, which results in
\begin{eqnarray}
\fl N_{\alpha}=\left[
2^{-2(n+l+1/2)}\pi^{1/2}R^{3}_{\alpha}\frac{(2n+2l)!}
{(n+l)!(n-l)!}\left\{1+\rho^{2}_{\alpha}\left(2n+l
-\frac{1}{2}\right)\right\}\right]
^{-1/2},
\end{eqnarray}
is obtained from the normalization condition
\begin{eqnarray}
\int^{\infty}_{0}\rmd^{3}\bi{x} \, u^{\dag}_{\alpha}(\bi{x})
u_{\alpha}(\bi{x})=1.
\end{eqnarray}

\newpage
\begin{table}
\caption{\label{tableI}Contributions of the individual diagrams
of Fig.2 to the axial charge $g_{A}$. Separate results for
the inclusion of ground (GS) and excited states (ES) in the quark
propagator are indicated.}
\begin{indented}
\item[]
\begin{tabular}{lr}
\br
&$g_A$\,\,\,\,\,\,\,\,\,\,\,\,\,\,\\
\mr
\bf{ GS quark propagator}\\
3q-core\\
\,\,\,\,\,\,\,\,\,LO&$1.25$\,\,\,\,\,\,\,\,\,\,\,\,\,\,
\,\,\,\,\,\,\,\,\,\,\\
\,\,\,\,\,\,\,\,\,NLO& $ -0.06$ \\
Counterterm & $-0.12$ \\
Exchange  & $ 0.23$ \\
Vertex correction & $ 0.01 $ \\
Self-energy & $ 0.19 $ \\
\mr
 GS contribution & $ 1.50 $ \\
\mr
\bf{ ES quark propagator}\\
NLO & $ -0.31 $ \\
Counterterm  & $ -0.25 $ \\
Vertex correction & $ 0.03 $\\
Self-energy  & $ 0.22 $\\
\mr
ES contribution & $ -0.31 $\\
\mr
Total(GS+ES)& \,$1.19 $\\
\mr
 Experiment\cite{Hagiwara}&$1.267 \pm 0.003$\,\,\,\\
\br
\end{tabular}
\end{indented}
\end{table}

\vspace*{-.5cm}
\begin{table}
\caption{\label{tableII}Comparison of the axial mass and the
axial radius between experimental values and the result
from the PCQM.}
\begin{indented}
\item[]
\begin{tabular}{@{}lcc}
\hline \hline &Model& \,\,\,\,\,\,\,\,\,\,\,\,\,\,\,\,\,\,
Experiment\,\,\,\,\,\,\,\,\,\,\,\,\,\,\,\,\,\,\,\\
\hline $M_A\,\,$(GeV) &$0.78$ &
$1.069\pm0.016$\cite{Bernard}\\
$\big<r^2_A\big>^{1/2}\,\,$(fm)\,\,\,\,\,\,\,\,\, &
$0.88$&
$0.639\pm0.010$\cite{Bernard}\\
\hline \hline
\end{tabular}
\end{indented}
\end{table}

\newpage

\begin{figure}
\begin{center}
\epsfig{figure=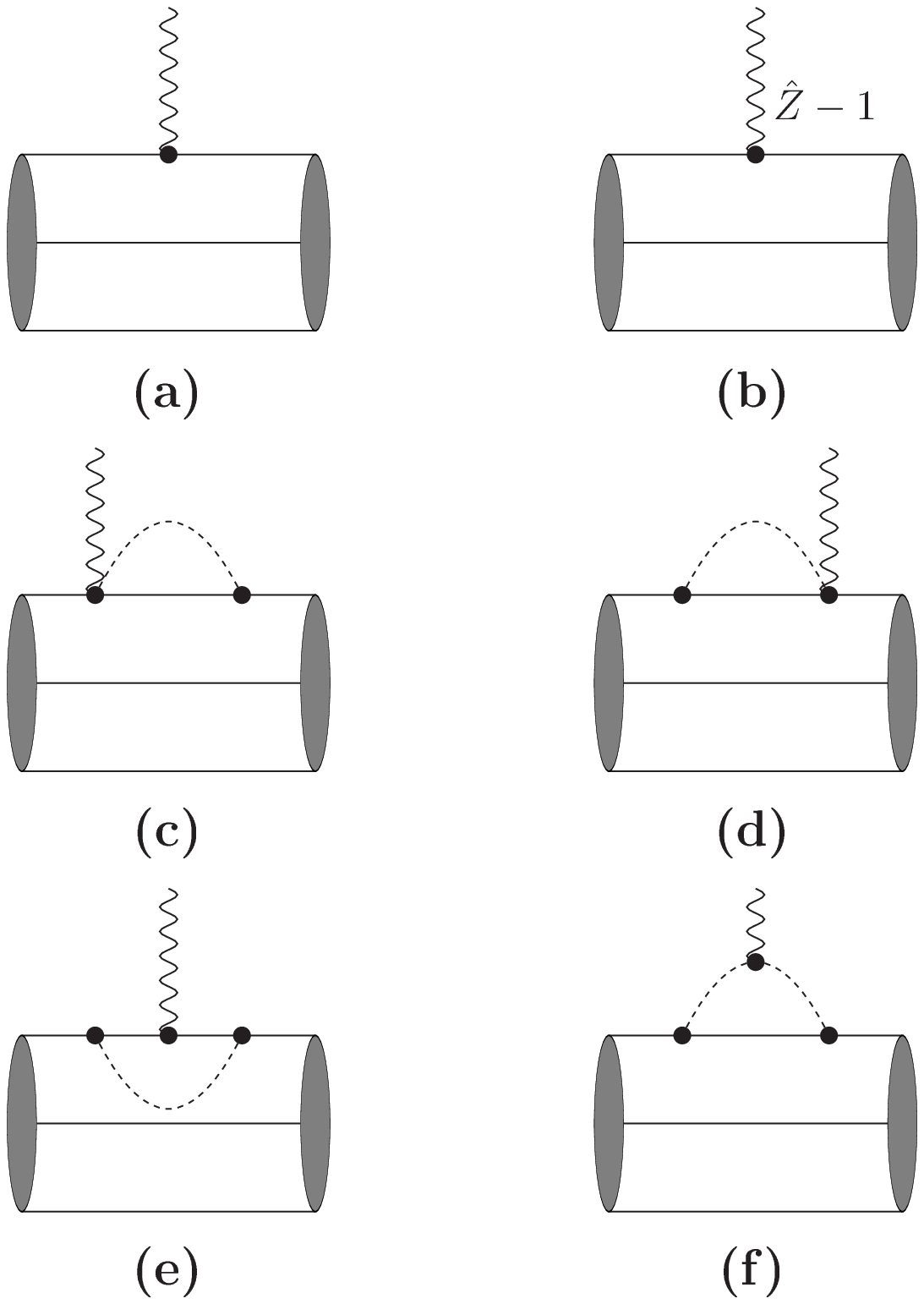,width=7cm}
\end{center}

\caption{\label{Fig1}Diagrams contributing to the nucleon charge:
triangle diagram (a), triangle counterterm diagram (b),
self-energy diagrams (c) and (d), vertex correction diagram~(e)
and meson-cloud diagram (f).}
\end{figure}

\newpage

\begin{figure}
\begin{center}
\epsfig{figure=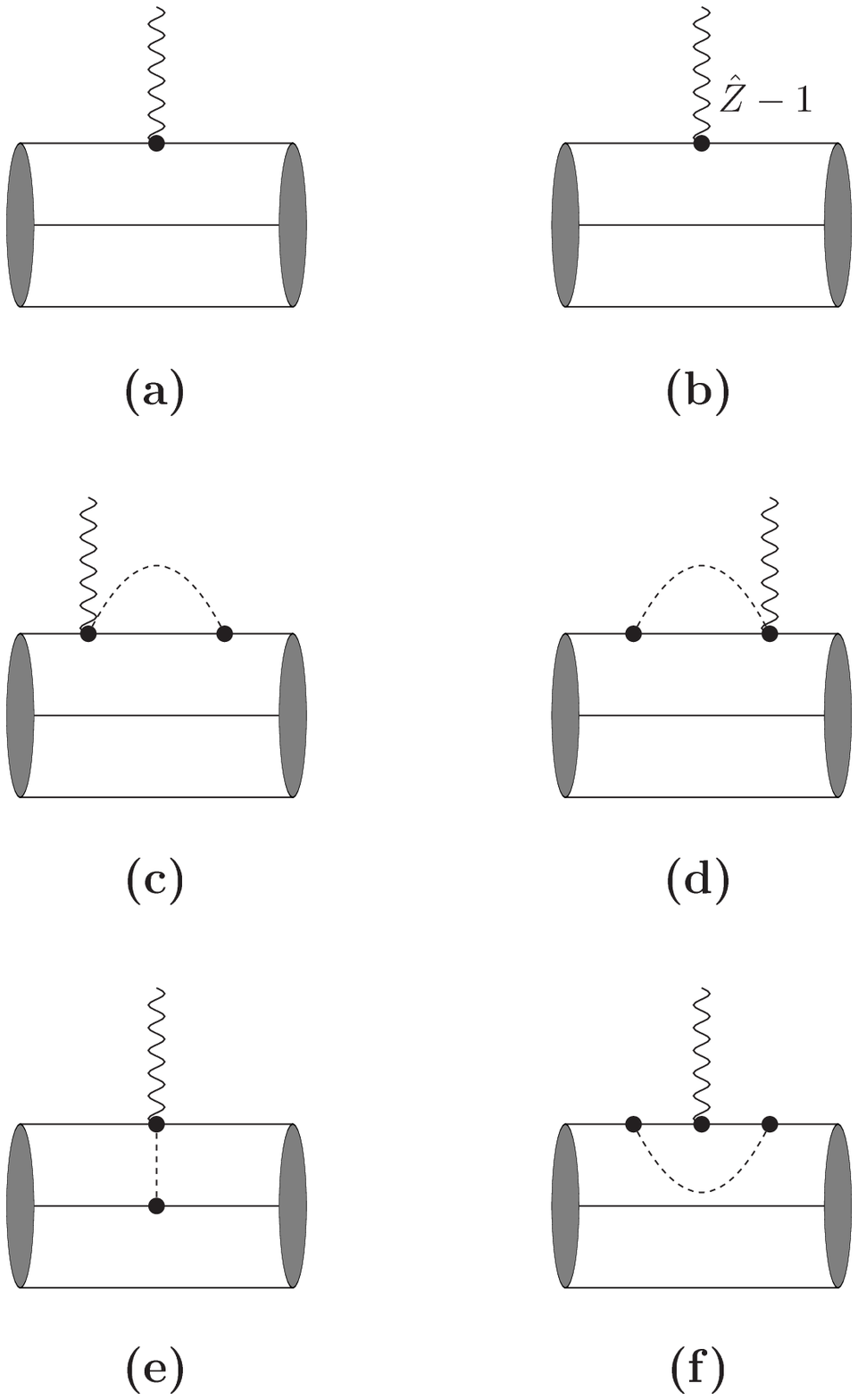,width=7cm}
\end{center}

\caption{\label{Fig2}Diagrams contributing to the axial form
factor of the nucleon : three-quark core (a), counterterm (b),
self-energy (c and d), exchange (e) and vertex correction (f).}
\end{figure}

\newpage

\begin{figure}
\begin{center}
\epsfig{figure=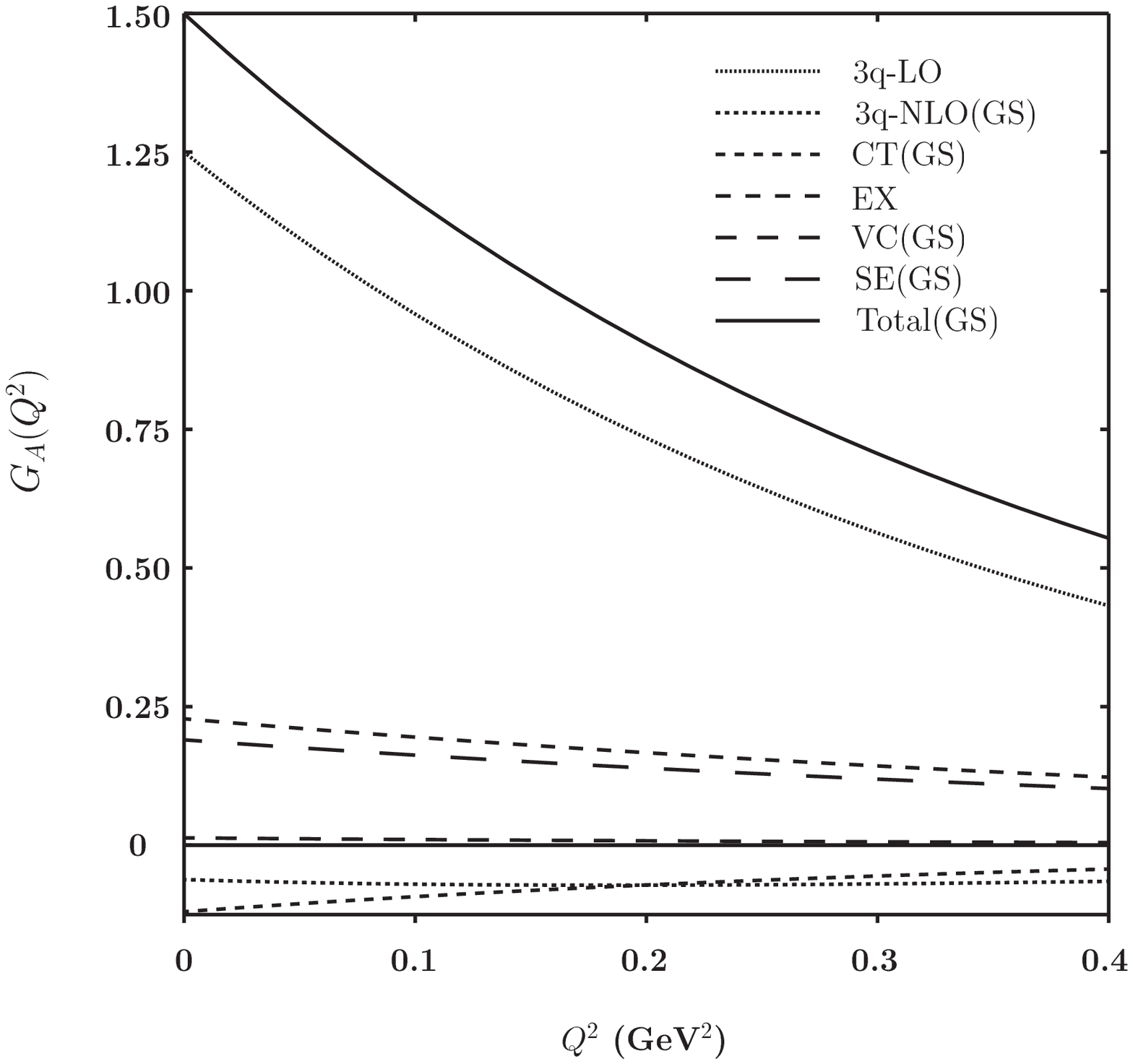,width=11cm}
\end{center}

\caption{\label{Fig3}Model result for the axial form factor of the
nucleon $G_{A}(Q^{2})$. The coherent contributions of the
different diagrams of Fig.2 are indicated when restricting to
the ground state(GS) quark propagator.}
\end{figure}

\newpage

\begin{figure}
\begin{center}
\epsfig{figure=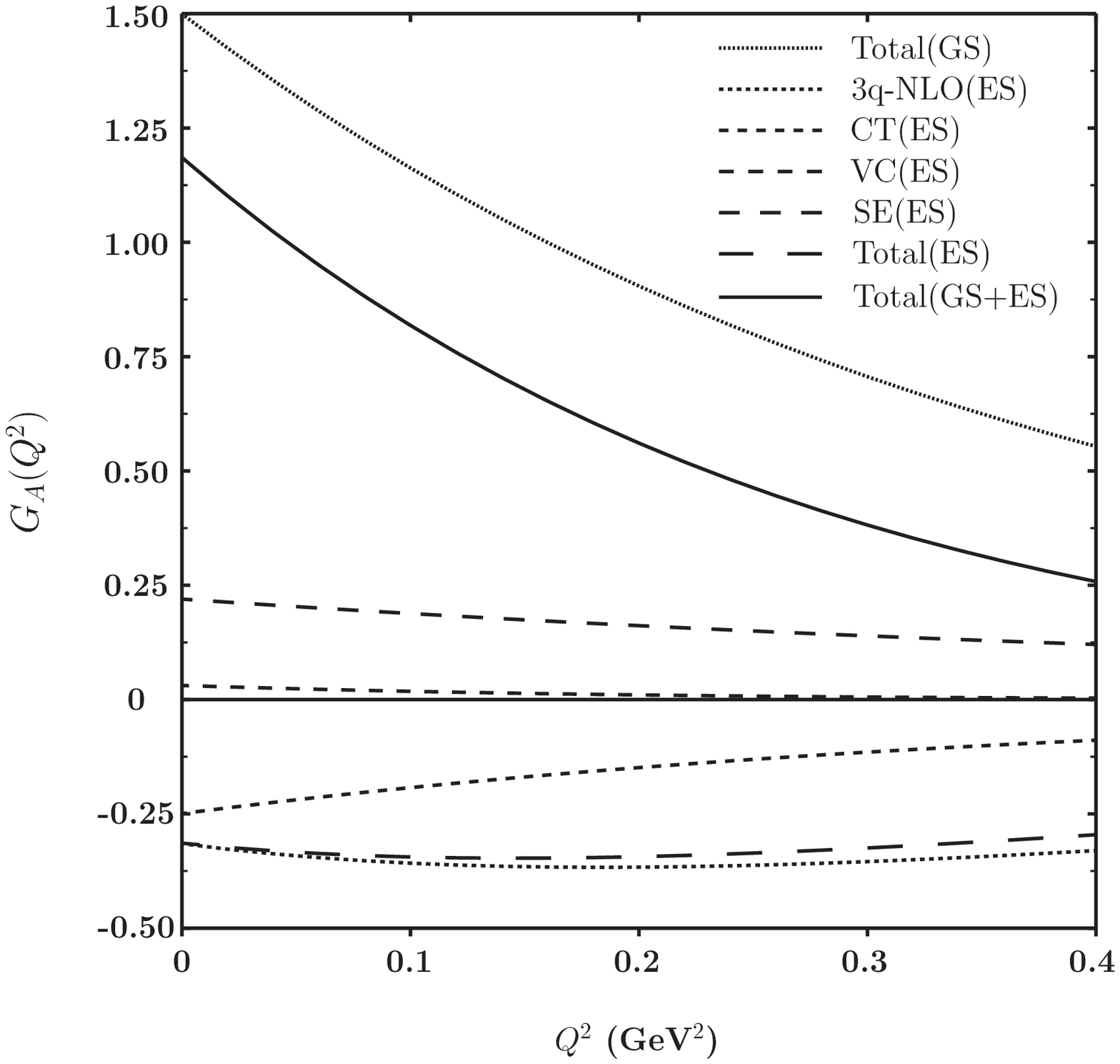,width=11cm}
\end{center}

\caption{\label{Fig4}Model result for the axial form factor of the
nucleon $G_{A}(Q^{2})$ when excited states are included in the
quark propagator. The full ground state result is contained in the
curve labelled by Total(GS). Excited state (ES) contributions of
the individual diagrams are indicated separately.}
\end{figure}

\newpage

\begin{figure}
\begin{center}
\epsfig{figure=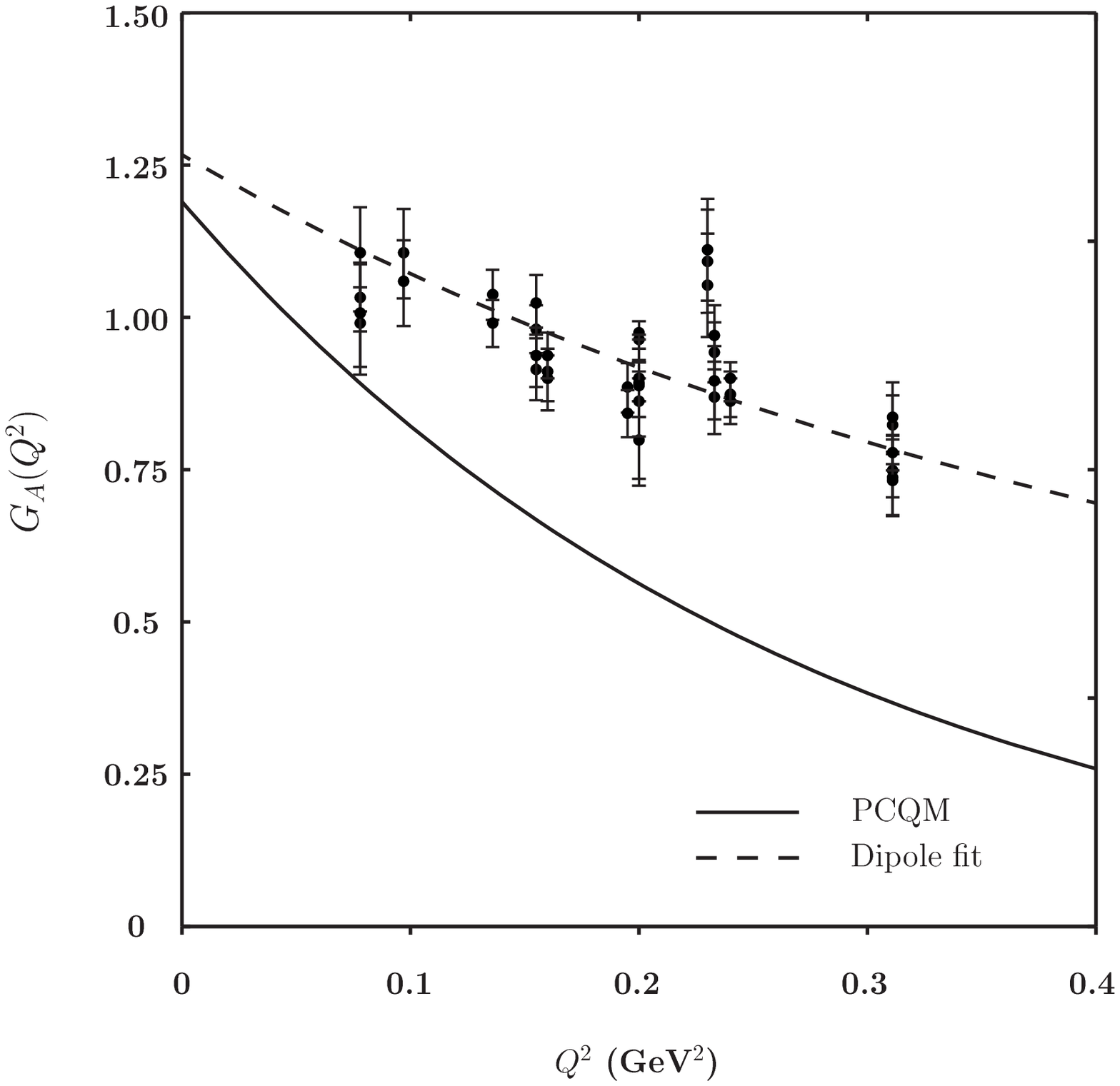,width=11cm}
\end{center}
\caption{\label{Fig5}The axial form factor of
the nucleon in the PCQM in comparison with a dipole fit (axial
mass $M_{A}=1.069$ GeV) and with experimental data. Data are taken
from references~\cite{Amaldi70}-\cite{Esaulov}.}
\end{figure}

\end{document}